\documentclass[journal=jpclcd,manuscript=article]{achemso}
\setkeys{acs}{articletitle=true}

\usepackage[version=3]{mhchem} 

\usepackage{amsmath}
\usepackage{amsfonts}
\usepackage{amssymb}
\usepackage{graphicx}
\usepackage{dcolumn}
\usepackage{bm}
\usepackage{subcaption}
\usepackage{algorithm} 
\usepackage{algpseudocode} 
\usepackage{xr}
\usepackage{hyperref}
\usepackage{braket}
\usepackage{xcolor}

\makeatletter
\newcommand*{\addFileDependency}[1]{
  \typeout{(#1)}
  \@addtofilelist{#1}
  \IfFileExists{#1}{}{\typeout{No file #1.}}
}
\makeatother

\usepackage[capitalize]{cleveref}
\crefname{figure}{Fig.}{Figs.}
\Crefname{figure}{Figure}{Figures}
\crefname{table}{Tab.}{Tabs.}
\Crefname{table}{Table}{Tables}
\crefname{equation}{Eq.}{Eqs.}
\Crefname{equation}{Equation}{Equations}
\crefname{section}{Sec.}{Secs.}
\Crefname{section}{Section}{Sections}
\usepackage{xcolor}

\usepackage{array}
\newcommand{\PreserveBackslash}[1]{\let\temp=\\#1\let\\=\temp}
\newcolumntype{C}[1]{>{\PreserveBackslash\centering}p{#1}}
\newcolumntype{R}[1]{>{\PreserveBackslash\raggedleft}p{#1}}
\newcolumntype{L}[1]{>{\PreserveBackslash\raggedright}p{#1}}

\usepackage{multirow}

\author{Anton Nyk{\"a}nen}
\affiliation{\fontsize{10pt}{10pt}\selectfont Algorithmiq Ltd., Kanavakatu 3C, FI-00160 Helsinki, Finland}

\author{Leander Thiessen}
\affiliation{\fontsize{10pt}{10pt}\selectfont Algorithmiq Ltd., Kanavakatu 3C, FI-00160 Helsinki, Finland}

\author{Elsi-Mari Borrelli}
\affiliation{\fontsize{10pt}{10pt}\selectfont Algorithmiq Ltd., Kanavakatu 3C, FI-00160 Helsinki, Finland}

\author{Vijay Krishna}
\affiliation{\fontsize{10pt}{10pt}\selectfont Department of Biomedical Engineering, Lerner Research Institute, Cleveland Clinic, Cleveland, 44195 OH, USA}
\alsoaffiliation{\fontsize{10pt}{10pt}\selectfont Department of Biomedical Engineering, Cleveland Clinic Lerner College of Medicine, Case Western Reserve University, Cleveland, 44106 OH, USA}

\author{Stefan Knecht}
\affiliation{\fontsize{10pt}{10pt}\selectfont Algorithmiq Ltd., Kanavakatu 3C, FI-00160 Helsinki, Finland}

\author{Fabijan Pavo\v{s}evi\'{c}}
\email{fabijan.pavosevic@algorithmiq.fi}
\affiliation{\fontsize{10pt}{10pt}\selectfont Algorithmiq Ltd., Kanavakatu 3C, FI-00160 Helsinki, Finland}

\title[]
  {$\Delta$ADAPT-VQE: Toward Accurate Calculation of Excitation Energies on Quantum Computers for BODIPY Molecules With Application in Photodynamic Therapy}

\begin{document}



\begin{tocentry}
\begin{figure}[H]
	\begin{center}
		\includegraphics[width=1.7in]{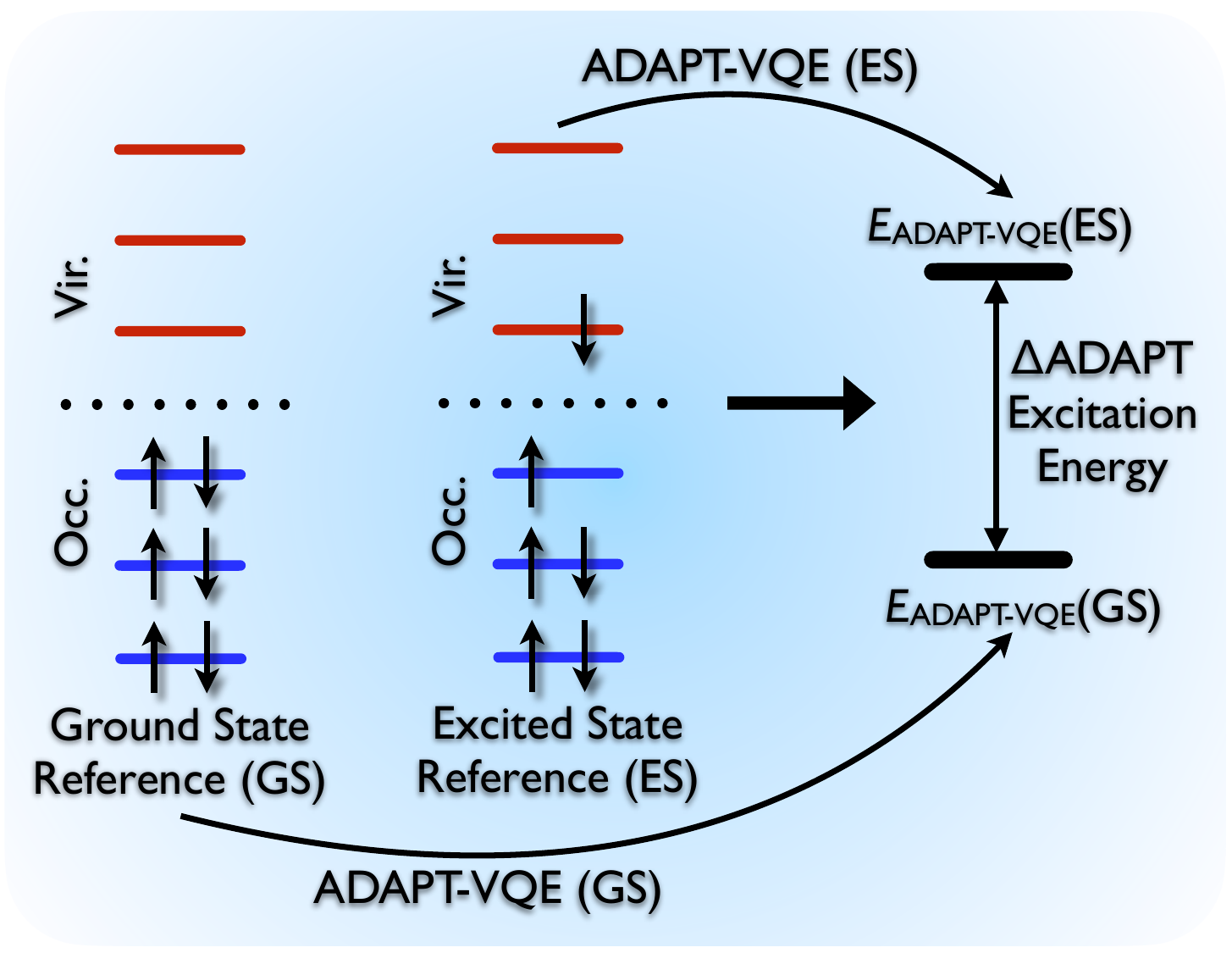}
	\end{center}
\end{figure}
\end{tocentry}

\begin{abstract}
Quantum chemistry simulations offer a cost-effective way for computational design of BODIPY photosensitizers with potential use in photodynamic therapy (PDT). However, accurate predictions of photophysical properties, such as excitation energies, pose a challenge for the popular time-dependent density functional theory (TDDFT) and equation-of-motion coupled cluster with singles and doubles (EOM-CCSD) methods. By contrast, reliable descriptions can be achieved by multi-reference quantum chemistry methods, though unfortunately, their computational cost grows exponentially with the number of correlated electrons. Alternatively, quantum computing holds a great potential for exact simulation of photophysical properties in a computationally more efficient way. To this end, we introduce the state-specific $\Delta$UCCSD-VQE (unitary coupled cluster with singles and doubles variational quantum eigensolver) and $\Delta$ADAPT-VQE methods in which the electronically excited state is calculated via a non-Aufbau electronic configuration. The accuracy and capability of the developed methods are  assessed against experimentally determined excitation energies for six BODIPY-derivatives. We show that the proposed methods predict accurate vertical excitation energies that are not only in good agreement with experimental reference data but also outperform popular quantum chemistry methods, such as TDDFT and EOM-CCSD. Spurred by its impressive performance and simplicity, we are confident that $\Delta$ADAPT will emerge as \textit{the} method of choice for guiding the rational design of photosensitizers for PDT and photocatalysis in the era of near-term quantum computing.

\end{abstract}

\maketitle

\section{Introduction}
Boron-dipyrromethene (also known as BODIPY) and its derivatives have emerged as an important class of organic fluorescent dyes characterized by high molecular absorption coefficients, high fluorescent quantum yields, and excellent thermal and photochemical stabilities~\cite{loudet2007bodipy}. These favorable photophysical properties can be further tuned by appropriate structural modifications of the BODIPY core~\cite{lu2014structural}, making them versatile compounds for medical imaging~\cite{yuan2013far}, labelling~\cite{gonccalves2009fluorescent}, photoelectrochemistry~\cite{nepomnyashchii2012electrochemistry}, artificial photosynthesis~\cite{bassan2021design}, optoelectronics~\cite{poddar2020recent}, and
photovoltaics~\cite{klfout2017bodipys}. BODIPYs can also be used in photodynamic therapy (PDT) as a photosensitizer (PS)~\cite{killoran2002synthesis,kamkaew2013bodipy}. PDT is an emerging procedure for noninvasive cancer treatment (including bladder, prostate, lung, breast, skin, and esophagus cancers) in which the PS upon light activation transfers electrons (Type I PDT mechanism) or energy (Type II PDT mechanism) to the surrounding environment to generate cytotoxic reactive species that cause cellular death (apoptosis and necrosis)~\cite{dolmans2003photodynamic,castano2006photodynamic}. The Type II PDT mechanism~\cite{derosa2002photosensitized} mainly occurs through the following three steps: (i) upon light absorption, the PS is excited to the first excited state S$_1$; (ii) a subsequent population of the first triplet (T$_1$) state via non-radiative intersystem crossing (ISC); (iii) energy transfer from PS in T$_1$ state to cellular triplet molecular oxygen to form reactive oxygen species (ROS). Unlike Type I, the Type II mechanism maintains the chemical form of the PS which allows for multiple therapeutic cycles during a single procedure. 

For BODIPYs to be useful in the PDT, they need to meet certain minimum criteria~\cite{turksoy2019photosensitization}, that are, non-toxic PS at the ground state S$_0$~\cite{breskey2013photodynamic}, a high and sharp absorption maximum within the tissue transparency window (750-900~nm),~\cite{feng2021perfecting} and a large S$_1$/T$_1$ spin-orbit coupling (SOC) to allow for an efficient formation of T$_1$ state with a high enough energy ($>$0.98~eV~\cite{herzberg1950spectra}) to generate ROS. To design functionalized BODIPYs with optimal photophysical properties mentioned above, quantum chemistry simulations play an important role~\cite{chibani2013revisiting}. From a theoretical standpoint, a computational aided design entails the calculation of singlet (S$_1$) and triplet (T$_1$) excitation energies (and potentially higher-excited states) which are important for the prediction of absorption spectra, fluorescent quantum yield~\cite{ou2020toward}, SOC~\cite{shaomin2015molecular}, and triplet lifetime. Because the S$_1$ is essential for many of these photophysical properties, the focus of the present work is primarily on the calculation of vertical excitation energies (S$_1$). Unfortunately, the most popular and efficient quantum chemistry method for excited states calculations, namely time-dependent density functional theory (TDDFT), is not able to accurately predict the photophysical properties of BODIPY dyes~\cite{chibani2012computation}. The predictive power of TDDFT strongly depends on the selected exchange-correlation functional exhibiting  errors in calculated S$_1$ excitation energies of 0.3-0.6~eV compared to experimentally determined values for a representative set of 17 chemically diverse BODIPY derivatives~\cite{momeni2015td}. Further improvements are obtained by employing various single-reference wave function approaches, such as equation-of-motion coupled cluster with singles and doubles (EOM-CCSD), second-order coupled cluster (CC2), scaled-opposite spin configuration interactions with singles and perturbative doubles (SOS-CIS(D)), and symmetry adapted cluster configuration interaction (SAC-CI), 
that yield on average an error on the order of 0.3~eV for calculated S$_1$ vertical excitation energies compared to experiment~\cite{chibani2014improving,momeni2015td,feldt2021assessment,berraud2019unveiling}. The most accurate vertical excitation energies are obtained by employing the multi-reference complete-active space second-order perturbation theory (CASPT2) method with errors less than 0.2~eV relative to experiment~\cite{momeni2015td}. Despite its success, the accuracy of CASPT2 strongly depends on the active space size~\cite{momeni2015td} and on the value of the shift parameter used to alleviate the intruder-state problem~\cite{roos1995multiconfigurational}. Moreover, since the number of configurations grows exponentially with increasing active space sizes, CASPT2 remains -- for practical purposes -- unsuitable for black-box applications to large molecular systems~\cite{Szalay2011}. 

By contrast, recent advances of quantum computing technology and quantum algorithms~\cite{bauer2020quantum,cao2019quantum} have opened a path to cope with the cost of exponential scaling. Therefore, quantum computing holds great promise for exact simulation of molecular processes in excited states involved in PDT. The most popular quantum computing algorithm for the currently available noisy intermediate-scale quantum (NISQ) devices~\cite{preskill2018quantum}, is the hybrid quantum-classical variational quantum eigensolver (VQE)~\cite{peruzzo2014variational}, which employs the quantum device for handling only classically intractable parts of the computation. It is used to find the ground state energy of a given molecular electronic Hamiltonian. The VQE algorithm has also been extended for excited state calculations, for example, by means of the quantum equation-of-motion (qEOM) approach~\cite{ollitrault2020quantum,asthana2023quantum} or a quantum subspace expansion (QSE) framework~\cite{mcclean2017hybrid} to name just a few. The overall accuracy and efficiency of any VQE \textit{ansatz} is determined by the choice of the wave function parametrization, and in its original implementation~\cite{peruzzo2014variational}, it exploited a fixed unitary coupled cluster with singles and doubles (UCCSD) ansatz~\cite{bartlett1989alternative,taube2006new} that demonstrates good performance for selected multi-reference problems~\cite{cooper2010benchmark,evangelista2011alternative,anand2022quantum}. However, a genuine implementation of UCCSD-VQE on NISQ devices requires very deep circuits, limiting its applicability only to the smallest chemical systems~\cite{peruzzo2014variational,o2016scalable,shen2017quantum,hempel2018quantum}. Alternatively, instead of employing a fixed UCCSD ansatz, the subsequently proposed  
ADAPT-VQE (Adaptive Derivative-Assembled Problem-Tailored)~\cite{grimsley2019adaptive} algorithm grows the ansatz iteratively, leading to significantly shallower circuits than those of UCCSD-VQE~\cite{grimsley2019adaptive}.

\begin{figure*}[ht!]
  \centering
  \includegraphics[width=6.5in]{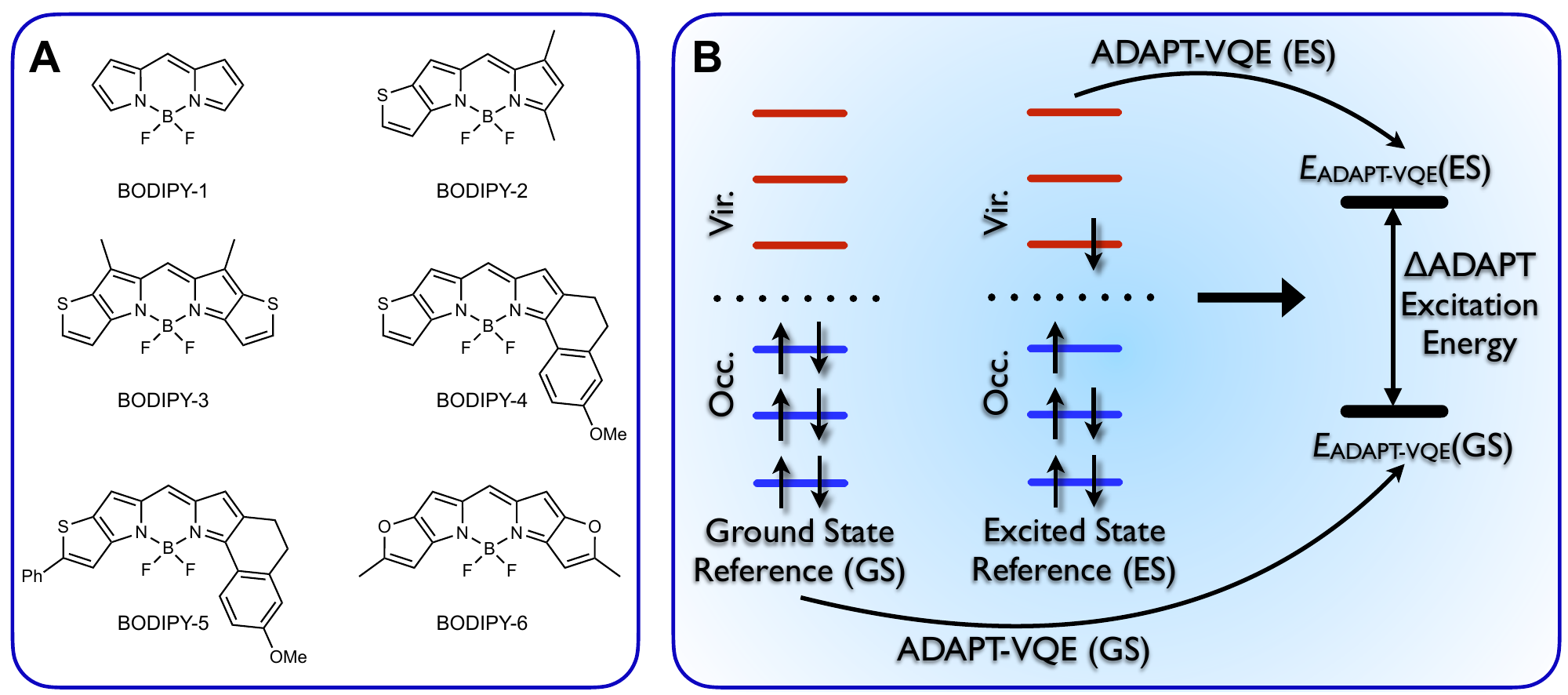}
  \caption{(A) Set of BODIPY structures considered in this work. (B) Schematic summary of the $\Delta$ADAPT-VQE method. }
  \label{fig:Figure1}
\end{figure*}

To lay the foundation for accurate simulations of vertical excitation energies for BODIPY molecules on the NISQ devices, in this work we introduce $\Delta$UCCSD-VQE and $\Delta$ADAPT-VQE, respectively, which are extensions of the classical $\Delta$SCF~\cite{ziegler1977calculation} (self-consistent field) and $\Delta$CCSD~\cite{lee2019excited,zheng2019performance} methods within a quantum computing framework. In the remainder of the text, we will refer to these methods simply as $\Delta$UCCSD and $\Delta$ADAPT. Both $\Delta$UCCSD and $\Delta$ADAPT are state-specific approaches in which the excited state is calculated via a non-Aufbau electronic configuration. We assess the performance and accuracy of $\Delta$UCCSD and $\Delta$ADAPT by comparing with  experimentally determined S$_1$ excitation energies of six BODIPY systems (see Fig.~\ref{fig:Figure1}A). Moreover, we discuss quantum resource estimates for an implementation of the two approaches on near-term quantum devices. The developments and numerical performance analysis emphasise the efficiency and accuracy of $\Delta$UCCSD and $\Delta$ADAPT geared toward the simulation of molecular excitation energies on the contemporary quantum computers. Moreover, this work lays the foundation for the development of other quantum-hardware efficient methods applicable to the calculation of excited-state energies and spectroscopic properties. 

\section{Theory}
The VQE~\cite{peruzzo2014variational} is a quantum computing algorithm used for finding the ground state energy of a molecular system defined by the electronic molecular Hamiltonian
\begin{equation}
    \label{eqn:Hamiltonian}
    \hat{H} = h^p_q a^q_p + \frac{1}{2}g^{pq}_{rs}a^{rs}_{pq}
\end{equation}
In this equation, $a^q_p = a_{q}^{\dagger}a_p$ and $a^{rs}_{pq} = a_{r}^{\dagger}a_{s}^{\dagger}a_qa_p$ are the one- and two-electron excitation operators (expressed in terms of fermionic creation/annihilation, $a^{\dagger}/a$, operators), respectively, whereas $h^p_q=\langle q |\hat{h}|p \rangle$ and $g^{pq}_{rs}=\langle rs|pq \rangle$ are one- and two-electron integrals in the spin-orbital basis, respectively. Througout this work, summation over repeating indices is implied. Moreover, indices denoted by $p,q,r,s,...$ correspond to general spin-orbitals, whereas indices denoted by $i,j,k,l,...$ and $a,b,c,d,...$ correspond to occupied and unoccupied (virtual) spin-orbitals, respectively. The ground state VQE energy, $E_{\text{VQE}}$, is obtained by minimizing the expectation value of the molecular Hamiltonian (defined in Eq.~\eqref{eqn:Hamiltonian}) 
\begin{equation}
    \label{eqn:VQE}
     E_{\text{VQE}}=\underset{\theta}{\text{min}} \langle\Psi(\theta)|\hat{H}|\Psi(\theta)\rangle
\end{equation}
with respect to parameters $\theta$ of some trial wave function $|\Psi(\theta)\rangle$.

In UCCSD-VQE, the wave function reads as
\begin{equation}
    \label{eqn:UCC}
    |\Psi_{\text{UCCSD}}\rangle=e^{\hat{T}-\hat{T}^\dagger}|0\rangle
\end{equation}
where $\hat{T}=\hat{T}_1+\hat{T}_2=\theta^{i}_{a}a_{i}^{a}+\frac{1}{4}\theta^{ij}_{ab}a_{ij}^{ab}$ is the cluster operator with singles and doubles and  $|0\rangle$ denotes a (single-determinantal) reference wave function, commonly coinciding with the Hartree-Fock (HF) determinant. UCCSD-VQE can be implemented with a polynomial number of one- and two-qubit quantum gates on NISQ devices, although in practice, even for small systems their number is too high, rendering this method impractical for larger molecular applications~\cite{peruzzo2014variational,o2016scalable,shen2017quantum,hempel2018quantum}. To push toward larger scale simulations, the ADAPT-VQE algorithm adaptively grows the wave function from a predefined set of the generalized fermionic pool of operators $\{a_p^q-a^p_q,a_{pq}^{rs}-a^{pq}_{rs}\}$~\cite{grimsley2019adaptive}. At each iteration of the ADAPT-VQE procedure, the operator with the highest electronic-energy gradient is added to augment the current trial wave function. Consequently, favorable features of the ADAPT-VQE algorithm are its accuracy paired with a resulting shallow circuit that allows for quantum chemistry simulations on NISQ devices.~\cite{rossmannek2023quantum}

As previously stated, the VQE algorithm is commonly used to find the ground state energy of a molecular system. In this work, we lift this limitation and introduce $\Delta$UCCSD and $\Delta$ADAPT, respectively, that enable us to locate a higher energy solution. Prediction of the excitation energies with the $\Delta$UCCSD or $\Delta$ADAPT methods, are obtained from two separate calculations; the ground state and the excited state energy calculations. The ground state energy is obtained by performing the usual UCCSD-VQE or ADAPT-VQE calculation with the Aufbau reference configuration. For obtaining the excited state energy, in the first step, the state-specific molecular orbitals (MO) of the non-Aufbau HF reference configuration are optimized via the SCF procedure (also known as $\Delta$SCF~\cite{ziegler1977calculation}). Such non-Aufbau reference configuration is constructed by placing one or more electrons in higher lying orbitals instead of the lowest lying orbitals (Aufbau). In the second step, these new set of orbitals are utilized to construct the state-specific molecular Hamiltonian defined in Eq.~\eqref{eqn:Hamiltonian}. Finally, using the Hamiltonian from the previous step within the VQE algorithm provides the excited state UCCSD-VQE or ADAPT-VQE energies. Once the ground and excited state UCCSD-VQE or ADAPT-VQE energies are obtained, their difference defines the $\Delta$UCCSD or $\Delta$ADAPT excitation energy, respectively. The schematic representation of the overall procedure is given in Fig.~\ref{fig:Figure1}B.

To reach quantitative predictions with quantum-chemical calculations, it is necessary to employ large and flexible basis sets. Unfortunately, such large basis sets drastically increase the number of qubits required for quantum simulations. The frozen natural orbital (FNO) approximation offers a robust path for reducing the number of qubits without sacrificing the accuracy~\cite{sosa1989selection}. Within the FNO approximation, the FNO orbitals (defined as the eigenvectors of the virtual-virtual block of the one-particle density matrix, $\gamma_a^b=\langle\Psi|a_a^b|\Psi\rangle$) are used to span the unoccupied orbital space. Moreover, their importance is ranked by their eigenvalues which correspond to the natural occupation numbers. Therefore, FNOs with larger natural occupation numbers will have a greater impact on the correlation energy than those with the smaller natural occupation numbers. Because the FNOs provide a more suitable basis for spanning the unoccupied orbitals compared to HF, for the same number of FNO and HF unoccupied orbitals, the FNO approach recovers significantly more correlation energy. In this work, the FNOs are obtained from the density matrix that employ the low-cost first-order Møller--Plesset (MP1) wave function, $|\Psi_{\text{MP1}}^{(1)}\rangle=\frac{1}{4}(g^{ij}_{ab}-g^{ij}_{ba})/(\epsilon_i+\epsilon_j-\epsilon_a-\epsilon_b)a_{ij}^{ab}|0\rangle$, where the $\epsilon$'s are the HF orbital energies. Lastly, for recovering the correlation energy that is outside of the reduced set of FNO orbitals, the missing correlation energy due to truncation is accounted by the following correction $\Delta E_{\text{MP2}}=E_{\text{MP2}}^{\text{MO}}-E_{\text{MP2}}^{\text{FNO}}$~\cite{neese2009efficient} that is added to the final energy. In this equation, $E_{\text{MP2}}^{\text{MO}}$ and $E_{\text{MP2}}^{\text{FNO}}$ are the energies calculated with the second-order Møller--Plesset (MP2) method in the untruncated MO basis and in the truncated FNO basis, respectively.

\section{Results and Discussion}
All of the reported results are obtained on the geometries of the six BODIPY systems compiled in Fig.~\ref{fig:Figure1} as obtained from Ref.~\citenum{berraud2019unveiling}. The TDDFT, EOM-CCSD, $\Delta$SCF, and $\Delta$CCSD calculations were performed with the \textsf{Q-Chem} quantum chemistry software~\cite{epifanovsky2021software} employing the cc-pVDZ basis set~\cite{dunning1989gaussian,woon1993gaussian}. The one-particle ($h$) and two-particle ($g$) molecular integrals that enter Hamiltonian defined by Eq.~\eqref{eqn:Hamiltonian} were obtained with the in-house modified version of \textsf{OpenFermion-QChem} library~\cite{OpenFermionQChem}. In all calculations, effects due to solvation are accounted for by the conductor-like polarizable continuum model (CPCM)~\cite{cossi2003energies}. The state-specific SCF procedure ($\Delta$SCF) utilizes the unrestricted orbitals and the maximum overlap method~\cite{gilbert2008self} for avoiding the variational collapse to the ground state. Fermionic operators were mapped into qubit space with the Jordan-Wigner mapper~\cite{jordan1993algebraic}. The $\Delta$UCCSD and $\Delta$ADAPT calculations are performed within \textsf{Qiskit Nature}~\cite{qiskit-nature} and Algorithmiq’s software framework \textsf{Aurora}, respectively, employing a noise-free quantum statevector simulator model. The reported $\Delta$UCCSD results are performed with (4,4), (6,6), and (8,8) active spaces, where the first value represents the number of electrons, and the second, the combined number of active occupied and natural (unoccupied) orbitals. The reported $\Delta$ADAPT results are performed with (6,6) and (8,8) active spaces. The reported ADAPT results utilize the procedure in which operators are selected from the operator pool consisting of majorana excitations (fermionic excitations split into invidiual Pauli strings, but without removing the Z-chains as usually done when using the Jordan-Wigner mapping)~\cite{tang2021qubit}. The number of two-qubit controlled NOT (CNOT) gates was estimated using the \textsf{Qiskit} transpile pass~\cite{Qiskit} for all-to-all qubit connectivity with optimization level set to 3.

\begin{figure*}[ht!]
  \centering
  \includegraphics[width=3.25in]{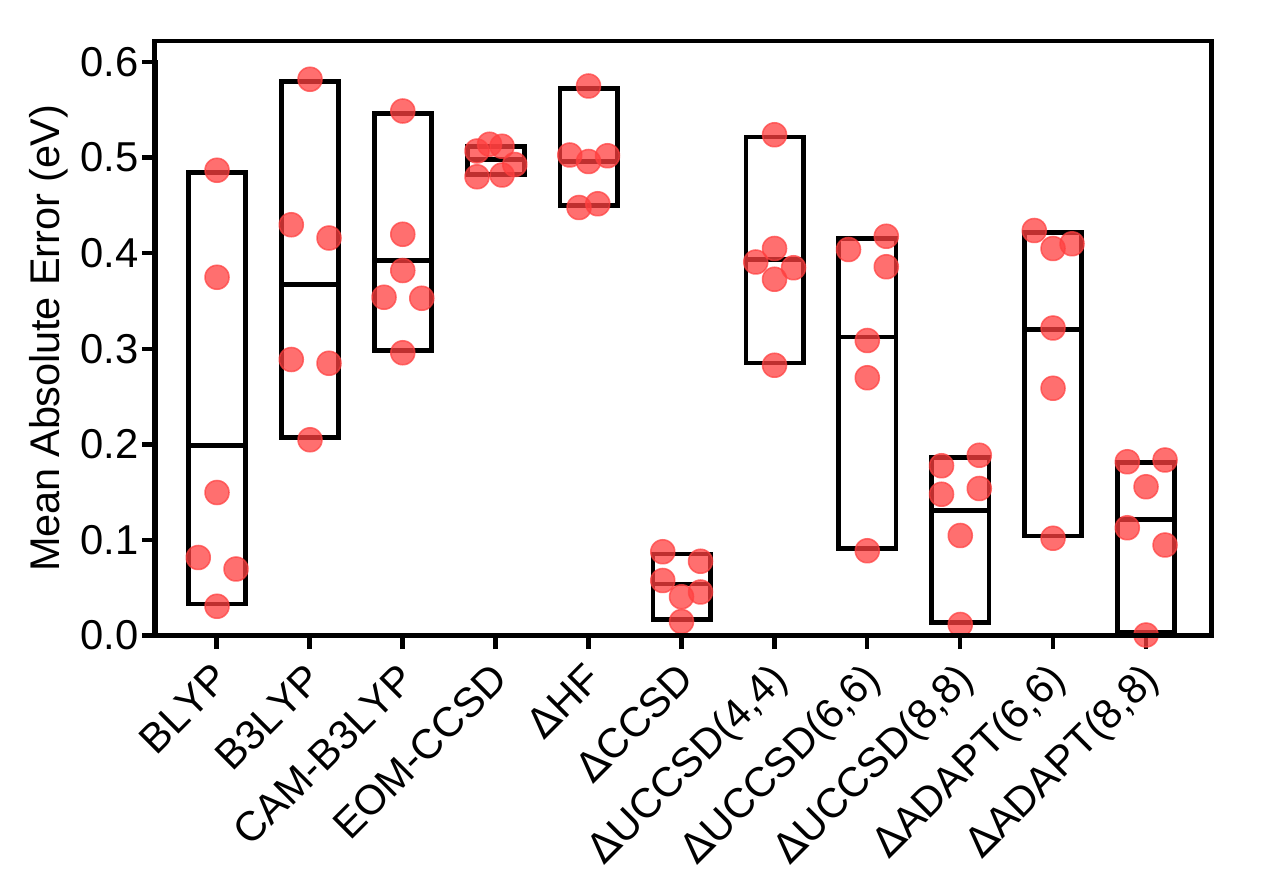}
  \caption{Individual absolute errors (AE) and mean absolute errors (MAE) of six BODIPY systems for different methods with respect to experimentally measured excitation energies (experimental data are obtained from Refs.~\citenum{arroyo2009smallest,jiang2012development,tanaka2013efficient,umezawa2009bright}). Calculations employ the cc-pVDZ basis set and solvent is treated with the CPCM implicit model.}
  \label{fig:Figure2}
\end{figure*}

To test the accuracy of the developed $\Delta$UCCSD and $\Delta$ADAPT quantum computing methods, we have calculated the vertical excitation energies ($\text{S}_0 \rightarrow \text{S}_1$) for a set of six BODIPY systems (Fig.~\ref{fig:Figure1}A) and compared the predicted excitation energies to the experimentally determined values (experimental data are obtained from Refs.~\citenum{arroyo2009smallest,jiang2012development,tanaka2013efficient,umezawa2009bright}). Figure~\ref{fig:Figure2} shows the individual absolute errors (AE) and mean absolute errors (MAE) of the calculated vertical excitation energies from experimental data, as obtained with different quantum chemical methods. Numerical values of the excitation energies calculated with different methods along with MAE and standard deviation (STDEV) are collected in Table~S1 of the Supporting Information (SI). The MAE for the TDDFT calculations with three different exchange-correlation functionals, BLYP~\cite{Parr88_785,Becke88_3098} (generalized gradient approximation), B3LYP~\cite{Parr88_785,Becke88_3098,beck1993density} (hybrid), and CAM-B3LYP~\cite{yanai2004new} (range-separated hybrid)
are 0.199~eV, 0.368~eV, and 0.392~eV, respectively. The same trend of performance for different functionals is also reported on a larger set of BODIPY systems~\cite{momeni2015td}. Overall, we observe that the TDDFT method with different functionals overestimate the experimental excitation energies by on average $\sim$0.3~eV which is in accordance with previous findings~\cite{momeni2015td,berraud2019unveiling}. While, BLYP exhibits the smallest MAE between these functionals, it has the largest deviation between the minimum and maximum value of the AE as shown in Fig.~\ref{fig:Figure2}. This is also supported by the largest value of STDEV as given in Table~S1. Among all studied methods herein, the largest MAE of 0.498~eV was observed for the EOM-CCSD method. However, unlike the TDDFT results, all of the EOM-CCSD excitation energy errors are closely clustered around the MAE. This is also evident from a very small STDEV of 0.015~eV and small deviation between the maximum and minimum AE. The $\Delta$HF method exhibits a very large discrepancy from experiments with MAE of 0.496~eV, which is mainly due to lack of electron correlation. Inclusion of electron correlation effects by moving to a CCSD ansatz within the $\Delta$CCSD method yields the best agreement with experiments for which MAE is only 0.054~eV. In passing we note that due to the non-unitary form of the ansatz, CCSD is unsuitable for implementation on quantum hardware.

Next, we discuss the performance of the quantum $\Delta$UCCSD and $\Delta$ADAPT methods proposed in this work. As shown in Fig.~\ref{fig:Figure2}, in case of $\Delta$UCCSD, the MAE systematically improves with increase of the active space size such that the $\Delta$UCCSD(8,8) method outperforms classical TDDFT and EOM-CCSD methods yielding MAE of 0.131~eV. Moreover, as shown in Table~S1, the $\Delta$UCCSD method provides nearly identical excitation energies to that of $\Delta$CCSD with the same active space size. This indicates that in the large active space limit, $\Delta$UCCSD will have very similar accuracy to $\Delta$CCSD. In addition, due to the variational nature of $\Delta$UCCSD, it is expected to outperform $\Delta$CCSD in situations where CCSD fails due to its nonvariational nature of the resulting wave function~\cite{cooper2010benchmark,evangelista2011alternative,kjonstad2017resolving,liu2021unitary,pavosevic2021polaritonic,kohn2022capabilities,culpitt2023unitary}. Finally, the $\Delta$ADAPT method shows nearly identical performance compared to $\Delta$UCCSD. We note that the reported $\Delta$ADAPT excitation energies are calculated as the energy difference between excited state and ground state wave functions that converged to within $1\mathrm{e}{-2}$~Hartree accuracy relative to the exact energy. Because of their unitary nature, both $\Delta$UCCSD and $\Delta$ADAPT are implementable on a quantum device. However, the implementation of individual states for $\Delta$UCCSD(6,6) and $\Delta$UCCSD(8,8) would require 9200 and 39568~CNOTs, respectively, rendering this simulations practically impossible on current near-term quantum devices. As already mentioned, we note that these estimates assume an all-to-all connectivity map of physical qubits as implemented in \textsf{Qiskit}. By contrast, $\Delta$ADAPT(6,6) and $\Delta$ADAPT(8,8) require on average (average between ground state and excited state ADAPT ansatz) \textit{merely} 70~CNOTs and 583~CNOTs, respectively (for more information regarding the CNOT count of each individual state see Table~S2). Importantly, the latter CNOT counts are well within contemporary quantum hardware capabilities and work on experimental hardware-based quantum simulations are ongoing in our laboratories. We would like to stress out that these active space sizes are tractable with the conventional multi-reference methods, however the simulations performed herein provides us with a valuable resources estimates and their scaling behavior as we increase the active space size, which is an important step for enabling simulations on the NISQ devices.

\section{Conclusion}
In this work, we have introduced the $\Delta$UCCSD and $\Delta$ADAPT methods for the calculation of excitation energies on noisy near-term quantum devices within a VQE algorithm formulation. The proposed approaches are non-Aufbau methods for which we employ in the first step non-Aufbau Hartree-Fock electronic references optimized by means of an SCF procedure~\cite{ziegler1977calculation}. Electronic correlation effects are subsequently accounted for by UCCSD-VQE and ADAPT-VQE, respectively. To  recover a significant portion of the total correlation energy, we additionally make use of the FNO approach. Moreover, solvation effects are taken into account with the CPCM implicit model. The developed methods were then employed to calculate vertical excitation energies ($\text{S}_0 \rightarrow \text{S}_1$) of six representative and chemically diverse BODIPY molecules with potential application in PDT. We compared computationally predicted excitation energies to corresponding experimentally determined values. The latter strikingly reveals that the proposed $\Delta$UCCSD and $\Delta$ADAPT approaches  outperform the popular classical TDDFT and EOM-CCSD methods. We also demonstrate that $\Delta$ADAPT can be implemented on NISQ devices by requiring only a modest number of CNOTs that is well within contemporary quantum hardware capabilities. 

The accuracy of both methods can be further improved by systematically increasing the active space size, thus approaching a $<$0.1~eV accuracy of predicted vertical excitation energy compared to experiment. Additional enhancement of the method compared to experiments (especially conducted in polar solvents that form hydrogen bonds with a solute) can be achieved by taking into account the explicit solvent effects instead of the implicit one, as employed herein, albeit at a higher computational cost due to statistical sampling~\cite{garcia2018simulation}. The use of spin-pure reference configurations instead of the unrestricted one could further improve the obtained results~\cite{schraivogel2021accuracy,kumar2022robust,tuckman2023aufbau}, and the work along those lines is in progress within our team.

In view of the accuracy and simplicity of $\Delta$ADAPT, we believe that this method will become \textit{the} method of choice for the calculation of molecular excitation energies on NISQ devices. Following the growing interest in the development of near-term quantum algorithms for excited state quantum chemistry applications, the $\Delta$ADAPT method will open new frontiers and guide an effective design of molecular photosensitizers equipped with the optimal properties for usage in photodynamic therapy~\cite{pham2021recent} and photocatalysis~\cite{prier2013visible}. Due to its variational nature, $\Delta$ADAPT will be particularly useful for NISQ simulations of strongly correlated transition metal-containing complexes that are at the core of the vast majority of  prominent photosensitizers~\cite{kim2024improved}. 

\begin{acknowledgement}
Work on “Quantum Computing for Photon-Drug Interactions in Cancer Prevention and Treatment” is supported by Wellcome Leap as part of the Q4Bio Program.
\end{acknowledgement}

\section*{Author contributions}
F.P. proposed the methods, wrote the code and the first draft. F.P. and V.K. prepared the figures. F.P. and A.N. obtained, evaluated, and analyzed the data. All authors participated in editing the manuscript.
\\

\noindent\textbf{Supporting Information Available}: The supporting information includes: experimental and calculated excitation energies for six BODIPY systems; number of two-qubit CNOT gates for ADAPT ansatz.
\\

\noindent\textbf{Conflict of interest}\\
The authors declare no conflict of interest.

\linespread{1}\selectfont
\bibliography{references}{}

\end{document}


\maketitle

\newpage
\section{Experimental and Calculated Excitation Energies for Six BODIPY Systems}

\begin{table}[]

\caption{Experimental excitation energies and calculated vertical excitation energies (eV) in solvent for six BODIPY molecules. Calculations employ the cc-pVDZ basis set and solvent is treated with the CPCM implicit model.}
\begin{tabular}{c|c c c c c c |c|c}
\hline
     & BOD.-1 & BOD.-2 & BOD.-3 & BOD.-4 & BOD.-5 & BOD.-6 & MAE\textsuperscript{a} & STDEV\textsuperscript{a} \\ \hline\hline
BLYP & 2.947 & 2.362 & 2.124 & 2.200 & 1.992 & 2.517 & 0.199 & 0.187 \\ \hline
B3LYP & 3.042 & 2.761 & 2.491 & 2.339 & 2.127 & 2.558 & 0.368 & 0.135 \\ \hline
CAM-B3LYP & 3.009 & 2.751 & 2.560 & 2.403 & 2.218 & 2.524 & 0.393 & 0.087 \\ \hline
EOM-CCSD & 2.967 & 2.845 & 2.688 & 2.562 & 2.402 & 2.635 & 0.498 & 0.015 \\ \hline
$\Delta$HF & 2.008 & 1.883 & 1.704 & 1.554 & 1.419 & 1.567 & 0.496 & 0.046 \\ \hline
$\Delta$CCSD & 2.445 & 2.273 & 2.128 & 2.091 & 1.968 & 2.054 & 0.054 & 0.026 \\ \hline
$\Delta$CCSD(4,4) & 2.054 & 1.968 & 1.815 & 1.767 & 1.539 & 1.617 & 0.392 & 0.078 \\ \hline
$\Delta$CCSD(6,6) & 2.371 & 2.063 & 1.819 & 1.630 & 1.610 & 1.738 & 0.313 & 0.124 \\ \hline
$\Delta$CCSD(8,8) & 2.612 & 2.243 & 2.026 & 1.861 & 1.924 & 1.993 & 0.127 & 0.071 \\ \hline
$\Delta$UCCSD(4,4) & 2.055 & 1.958 & 1.815 & 1.767 & 1.537 & 1.618 & 0.394 & 0.077 \\ \hline
$\Delta$UCCSD(6,6) & 2.371 & 2.061 & 1.820 & 1.632 & 1.613 & 1.738 & 0.313 & 0.124 \\ \hline
$\Delta$UCCSD(8,8) & 2.614 & 2.226 & 2.028 & 1.861 & 1.934 & 1.994 & 0.131 & 0.065 \\ \hline
$\Delta$ADAPT(6,6) & 2.358 & 2.072 & 1.782 & 1.645 & 1.600 & 1.732 & 0.320 & 0.125 \\ \hline
$\Delta$ADAPT(8,8) & 2.573 & 2.236 & 2.024 & 1.866 & 1.921 & 1.986 & 0.122 & 0.069 \\ \hline
Experiment & 2.460\textsuperscript{b} & 2.331\textsuperscript{c} & 2.206\textsuperscript{d} & 2.050\textsuperscript{e} & 1.922\textsuperscript{e} & 2.142\textsuperscript{f} & - & - \\ \hline\hline
\end{tabular}

\raggedright \textsuperscript{a}\small Mean absolute error (MAE) and standard deviation (STDEV) of calculated excitation energies with respect to experimental data. \\

\raggedright \textsuperscript{b}\small Experimental measurements are done in cyclohexane~\cite{arroyo2009smallest}. \\

\raggedright \textsuperscript{c}\small Experimental measurements are done in THF~\cite{jiang2012development}. \\

\raggedright \textsuperscript{d}\small Experimental measurements are done in  dichloromethane~\cite{tanaka2013efficient}. \\

\raggedright \textsuperscript{e}\small Experimental measurements are done in chloroform~\cite{jiang2012development}. \\

\raggedright \textsuperscript{f}\small Experimental measurements are done in chloroform~\cite{umezawa2009bright}. \\

\label{table:tableS1}
\end{table}

\newpage
\section{Number of Two-Qubit CNOT Gates for ADAPT Ansatz}

\begin{table}[]

\caption{Number of CNOTs for the ground and excited state ADAPT ansatze. Both ground and excited state ADAPT wave functions are converged to within $1\mathrm{e}{-2}$~Hartree accuracy relative to the exact energy.}

\begin{tabular}{ c|cc|cc }
\hline
 & \multicolumn{2}{c|}{ADAPT(6,6)}    & \multicolumn{2}{c }{ADAPT(8,8)}    \\ \hline
 & \multicolumn{1}{c|}{Ground State} & Excited State & \multicolumn{1}{c|}{Ground State} & Excited State \\ \hline\hline
BOD.1 & \multicolumn{1}{c|}{162} & 123 & \multicolumn{1}{c|}{896} & 1035 \\ \hline
BOD.2 & \multicolumn{1}{c|}{90} & 38 & \multicolumn{1}{c|}{766} & 719 \\ \hline
BOD.3 & \multicolumn{1}{c|}{16} & 15 & \multicolumn{1}{c|}{539} & 407 \\ \hline
BOD.4 & \multicolumn{1}{c|}{87} & 109 & \multicolumn{1}{c|}{458} & 300 \\ \hline
BOD.5 & \multicolumn{1}{c|}{62} & 29 & \multicolumn{1}{c|}{465} & 65 \\ \hline
BOD.6 & \multicolumn{1}{c|}{44} & 58 & \multicolumn{1}{c|}{780} & 556 \\ \hline
Average & \multicolumn{1}{c|}{77} & 62 & \multicolumn{1}{c|}{651} & 514 \\ \hline\hline
\end{tabular}
\end{table}

\linespread{1}\selectfont
\bibliography{references}{}